# Inverse Modeling of Climate Responses of Monumental Buildings

R.P. Kramer, A.W.M. van Schijndel & H.L. Schellen
Eindhoven University of Technology

**ABSTRACT**

The indoor climate conditions of monumental buildings are very important for the conservation of these objects. Simplified models with physical meaning are desired that are capable of simulating temperature and relative humidity. In this paper we research state-space models as methodology for the inverse modeling of climate responses of unheated monumental buildings. It is concluded that this approach is very promising for obtaining physical models and parameters of indoor climate responses. Furthermore state space models can be simulated very efficiently: the simulation duration time of a 100 year hourly based period take less than a second on an ordinary computer.

**INTRODUCTION**

Effects of climate change on ecosystems and on the global economy have been researched intensively during the past decades but almost nothing is known about the influence to our cultural heritage. Although these historical monuments are exposed to extensive loads caused by stampedes of visitors, there are many other factors deteriorating World Heritage Sites. The impacts of climate change are a long-term and substantial menace to the sites. Lots of monumental buildings are used as museum or storage for paintworks, books and artefacts. The indoor climate conditions of monumental buildings are very important for the conservation of these objects (ASHRAE, 2007). The influence of the changing climate on the indoor climate of monumental buildings is unknown. It is impossible to prepare adequately for the future, by anticipating on this change and adapting the installations, because of this lack of knowledge (CIBSE, 2005). The result is that the conservation of the buildings and the collections are at risk. Furthermore, the worldwide energy problem also affects these monumental buildings. Due to the ancient building techniques used, which result in huge transmission and infiltration losses, the energy consumption of these buildings is high. Three problems can be identified with respect to the current research methods: Firstly, due to the long simulation period (hundred years with time step 1h), combined with detailed physical models, the simulation run time is long. Secondly, the detailed modeling of the buildings itself requires much effort: the monumental buildings are old and protected. Therefore, blueprints are hard to find and; destructive methods to obtain building material properties are not allowed. Thirdly, the used modeling approach does neither facilitate in an easy characterization of the building nor in an easy characterization of the energy performance. A simplified model with physical meaning is desired which is capable of simulating both temperature and relative humidity. The objective is the successful application of inverse modeling on a simplified thermal and hygric building model, in order to determine the parameters with physical meaning. The simplified model is needed for the prediction of the indoor temperature and the indoor relative humidity and for characterization of the building parameters and the energy performance. The paper is organised as follows. We start with a literature study on simplified models. The methodology of the inverse modeling development for climate responses is presented and applied for a group of four unheated monumental buildings in The Netherlands and Belgium. The approach is evaluated and conclusions are drawn.

**LITERATURE**

Due to the increase of computational power, the attention for simplified models has decreased. However, through the years it became clear that simplified models have benefits over complex models (Wang & Chen, 2001), (Mathews, et al., 1994): user friendliness, straight forward, fast calculation. The response factor method and lumped capacitance method are suitable for simplified modeling. More recently, linear parametric models and neural network models are used for simplified models. Neural network models, e.g. (Mustafaraj, et al., 2011), can be classified as black box models. The parameters have no direct physical meaning, but the output is generated by the hidden layers (black box) from the input. Some models are referred to as gray box models. An example in the field of simplified building models is the use of linear parametric models (Mustafaraj, et al., 2010).The linear model itself is a black box model, but the parameters can be determined using physical data (Jimenez, et al., 2008). Some researchers stress out the importance of simplified models with physical meaning (Kopecký, 2011), so called white box models. The lumped capacitance model can be classified as a white box model. Another advantage of this approach is the representation of building elements using R (resistance) and C (capacitance), according to the electrical analogy, which makes a graphical representation of the model possible. Most of the simplified building models are based on this approach. There are three approaches to create a simplified model: (1) Create a detailed comprehensive model from known building properties and perform afterwards a model order reduction technique, e.g. (Gouda, et al., 2002); (2) Create directly a simplified model from building properties, e.g. (Nielsen, 2005); (3) Create a simplified model and identify the parameter values with an inverse modeling technique (Balan, et al., 2011). Technique 1 is obviously the most labor intensive: building a detailed model and simplifying it afterwards. Detailed construction properties need to be available together with a methodology for simplifying an existing model. The lumped capacitance model can be used for this model order reduction (Mathews, et al., 1994) and neural network models can be used to filter out unimportant parameters (Mustafaraj, et al., 2011), called pruning. Technique 2 is faster, but a validated methodology should be known how to identify the parameters. Therefore, it is difficult to achieve good results with this technique. (Fraisse, et al., 2002) has demonstrated a methodology how to incorporate multiple walls into one single order model. Technique 3 is not labor intensive and identification of the model parameters is done by an optimization algorithm. This technique can be used with the lumped capacitance model (Wang & Xu, 2006), neural network model, e.g. (Mustafaraj, et al., 2011), and linear parametric model, e.g. (Moreno, et al., 2007). Technique 3 is used for the work in this paper.

**INVERSE MODEL DEVELOPMENT FOR CLIMATE RESPONSES**

A linear model is often sufficient to accurately describe the system dynamics and, in most cases, one should first try to fit linear models (Ljung, 1999). State Space models are Linear Time Invariant (LTI) models. State Space models are models that use state variables to describe the system dynamics by a set of first order differential equations. To understand the concept of State Space, think of the system as spanning a space where the axes (i.e. dimensions, i.e. orders) represent the state variables. Even if some of the systems differential equations are higher order, they should be converted to multiple first order equations. The state of the system can be represented as a vector within that space. The system of first order differential equations can be represented according to:

$$\dot{x}(t) = Ax(t) + Bu(t)$$
$$y(t) = Cx(t) + Du(t)$$

The first part of equation (1) is known as *state equation* where **x**(t) is the state vector and **u**(t) is the input vector. The second equation is referred to as the output equation. **A** is the state matrix, **B** is the input matrix, **C** the output matrix and **D** the direct transition matrix. The main advantage is that the calculation speed is very high, especially compared to solving the differential equations. To show this advantage, a first order RC-network is simulated as a State Space model and with the ode23 routine for different simulation periods. The results are shown in Table 1.

*Table 1: calculation time of State-Space and ode23 for different periods.*

|  | 1month time [s] | 1 year time [s] | 10 year time [s] | 100 year time [s] |
|---|---|---|---|---|
| ODE23 | 5 | 89 | - | - |
| State-Space | 0.016 | 0.016 | 0.050 | 0.45 |

The results show the huge advantage of the State Space model regarding simulation time. While the simulation time increases almost linearly when using the ode23 routine to solve the differential equation, the simulation time of the State Space model is less predictable. However, especially if the task requires a repeated simulation for a long period, e.g. hundred years, the State Space model has a very limited calculation time of half a second on an ordinary computer.

We proceed with the thermal and hygric modeling. Due to space limitation we summarize the work of Kramer (2012). He investigated several thermal models including solar irradiation. Also several hygric models have been developed and tested. In this paper we present the optimal models according to Kramer (2012).

The thermal model is shown in Figure 1. There are thermal capacitances for the interior (Cint), indoor air (Ci) and envelope (Cw). The sun irradiation is connected to Cint (interior) and Ci (indoor air). Thermal resistances represent ventilation (1/Gfast), wall to external air (1/Gw), wall to indoor air (1/Gi) and indoor air to interior constructions (1/Gint).

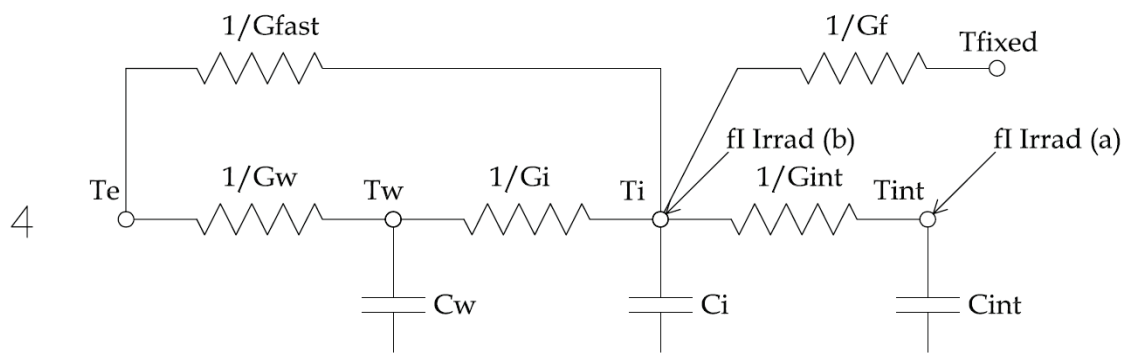

*Figure 1. The thermal network*

The ODEs are,

$$C_w \frac{dT_w}{dt} = G_w(T_e - T_w) - G_i(T_w - T_i)$$

$$C_i \frac{dT_i}{dt} = G_i(T_w - T_i) - G_f(T_i - T_f) - G_{int}(T_i - T_{int}) + G_{fast}(T_e - T_i) + \vec{fI} \cdot \vec{Irrad}(b)$$

$$C_{int} \frac{dT_{int}}{dt} = G_{int}(T_i - T_{int}) + \vec{fI} \cdot \vec{Irrad}(a)$$

The State Space matrices are,

$$A = \begin{bmatrix} \frac{-G_w-G_i}{C_w} & \frac{G_i}{C_w} & 0 \\ \frac{G_i}{C_i} & \frac{-G_i-G_f-G_{int}-G_{fast}}{C_i} & \frac{G_{int}}{C_i} \\ 0 & \frac{G_{int}}{C_{int}} & \frac{-G_{int}}{C_{int}} \end{bmatrix}$$

$$B = \begin{bmatrix} \frac{G_w}{C_w} & 0 & 0 & 0 & 0 & 0 \\ \frac{G_{fast}}{C_i} & \frac{fI_1}{C_i} & \frac{fI_2}{C_i} & \frac{fI_3}{C_i} & \frac{fI_4}{C_i} & \frac{G_f}{C_i} \\ 0 & 0 & 0 & 0 & 0 & 0 \end{bmatrix}$$

$$C = \begin{bmatrix} 0 & 0 & 0 \\ 0 & 1 & 0 \\ 0 & 0 & 0 \end{bmatrix} \qquad D = \begin{bmatrix} 0 & 0 & 0 & 0 & 0 & 0 \\ 0 & 0 & 0 & 0 & 0 & 0 \\ 0 & 0 & 0 & 0 & 0 & 0 \end{bmatrix}$$

The hygric model is shown in Figure 2. The hygric capacitances and resistances represent are analog to the thermal network, but now for vapour pressures instead of temperatures.

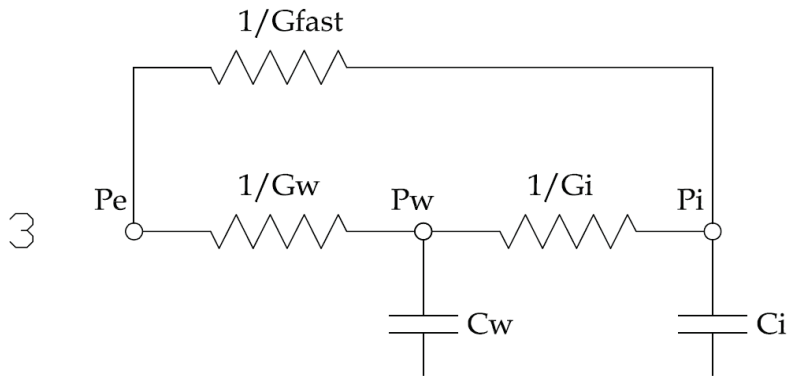

Figure 2. The hygric model

The ODEs are

$$C_w \frac{dP_w}{dt} = G_w(P_e - P_w) - G_i(P_w - P_i)$$

$$C_i \frac{dP_i}{dt} = G_i(P_w - P_i) + G_{fast}(P_e - P_i)$$

The State Space matrices are,

$$A = \begin{bmatrix} \frac{-G_w-G_i}{C_w} & \frac{G_i}{C_w} \\ \frac{G_i}{C_i} & \frac{-G_i-G_{fast}}{C_i} \end{bmatrix} \quad B = \begin{bmatrix} \frac{G_w}{C_w} \\ \frac{G_{fast}}{C_i} \end{bmatrix} \quad C = \begin{bmatrix} 0 & 0 \\ 0 & 1 \end{bmatrix} \quad D = \begin{bmatrix} 0 \\ 0 \end{bmatrix}$$

This completes the modeling part. We proceed with the inverse modeling methodology.

Inverse modeling is the inverse of traditional modeling. In traditional modeling, the system is known and the output is unknown. By modeling the system, the output can be simulated. In inverse modeling, the output is known (e.g. measured), but little is known about the system. The objective is to identify the parameter values of the model by repeatedly trying different parameter values and comparing the simulated output with the measured output. The goal is to minimize the simulation error, formulated as an objective function (e.g. summed squared error). The process of finding the parameter set which minimizes the objective function is called optimization. If the solution space includes multiple minima, the goal is to find the global minimum, called global optimization. These solvers are all very different, each having its advantages and disadvantages. One important aspect is whether a solver is gradient based or gradient free. Gradient based solvers are the most efficient in finding quickly a minimum. However, the solution space should be smooth and continuous. If not, the solver fails. The second aspect is whether the solver handles constraints or it is only intended for unconstrained problems. The inverse problem in this research is a typical example of a constrained problem, since all variables are not allowed to be negative. Moreover, constraining the problem helps in finding the global minimum since it scales down the solution space. The third aspect is whether a solver is deterministic or stochastic: all solvers are deterministic except for the Genetic Algorithm. To maximize the speed of the optimization process, all calculations which do not need to be repeated are executed in the initialization step: preparing climate data, include measured data, set constraints, set initial values and assign a function handle for the optimization algorithm. Then the optimization algorithm determines the parameter set, the first time it's the initial value vector, and passes the parameter set to the function file: the function file includes the model and simulates the model with the given parameter set and calculates the objective function by comparing the simulated output with the measured data. The objective function is passed to the Global Optimization Algorithm which calculates the new parameter set which is likely to minimize the objective function.

**RESULTS FOR A GROUP OF UNHEATED MONUMENTAL BUILDINGS**
An important validation method is the performance assessment of the developed models on multiple other buildings: By fitting both the thermal and hygric model to these buildings, their general performance is tested. General applicability is an important aspect in this study. Moreover, it is important to chose a strategic set of buildings and rooms to gain maximum added value of this validation method: included in the set of buildings are (i) Room which is surrounded by a water canal; (ii) Room with much thermal mass and hygric mass (iii) Room with significant sun irradiation (iv) Room with sun irradiation on roof, but no windows (v) Rooms with ground contact and no sun irradiation. The physical aspects of the buildings and rooms and the argumentation why they are included, is explained in the next Section. The performance per room is visualized in graphs and explained in word. The performance of the rooms is compared by three performance criteria (MSE, MAE & Goodness of Fit) at the end of the section. Figure 3 presents the monumental buildings involved in the study.

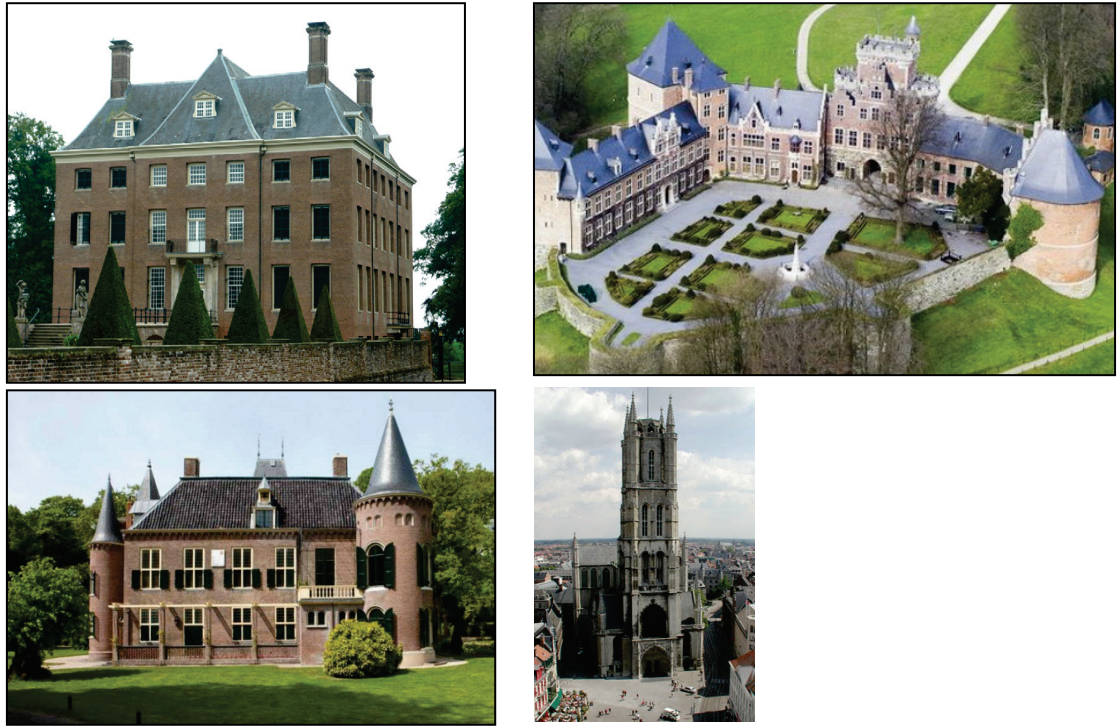

*Figure 3. The Monumental buildings involved in this study. Top Left: Castle of Amerongen, Top Right: Castle of Gaasbeek, Bottom Left: Castle Keukenhof, Bottom Right: St. Baafs Cathedral*

*Castle of Amerongen: Washing room*
The washing room is situated in the basement of the castle of Amerongen. The castle is surrounded by a canal: the external walls of the washing room are in direct contact with water. Moreover, the basement has been flooded multiple times. The performance of the thermal model is shown in Figure 4 (top). The model performs good over the seasons resulting in a good fit. The detail at the right side of the figure shows that the delicate fluctuations of the measured signal are not reproduced by the model, which is a good thing: the sensors have an accuracy of ± 0.5°C meaning that the observed fluctuations are cover by the uncertainty of the measurement. The reason of these delicate fluctuations is unclear and considered to be unimportant for the performance assessment of the model. The performance of hygric model 3 is shown in Figure 4 (bottom). There seems to be a problem analogous to the thermal model which has been observed earlier: the signal lies locally above, but mostly below the measured signal. The result is that the model can't be fitted accurately. A new parameter is introduced for the hygric model, which is analogous to the thermal model: a node with a fixed vapour pressure. The performance of this hygric model with fixed vapour pressure is shown in Figure 4 (middle). The fixed vapour pressure improves the fit significantly. The necessity of such a fixed vapour pressure suggest the existence of a vapour source. Physically, the external wall which is connected to a canal is analogous to the thermal situation with ground contact.

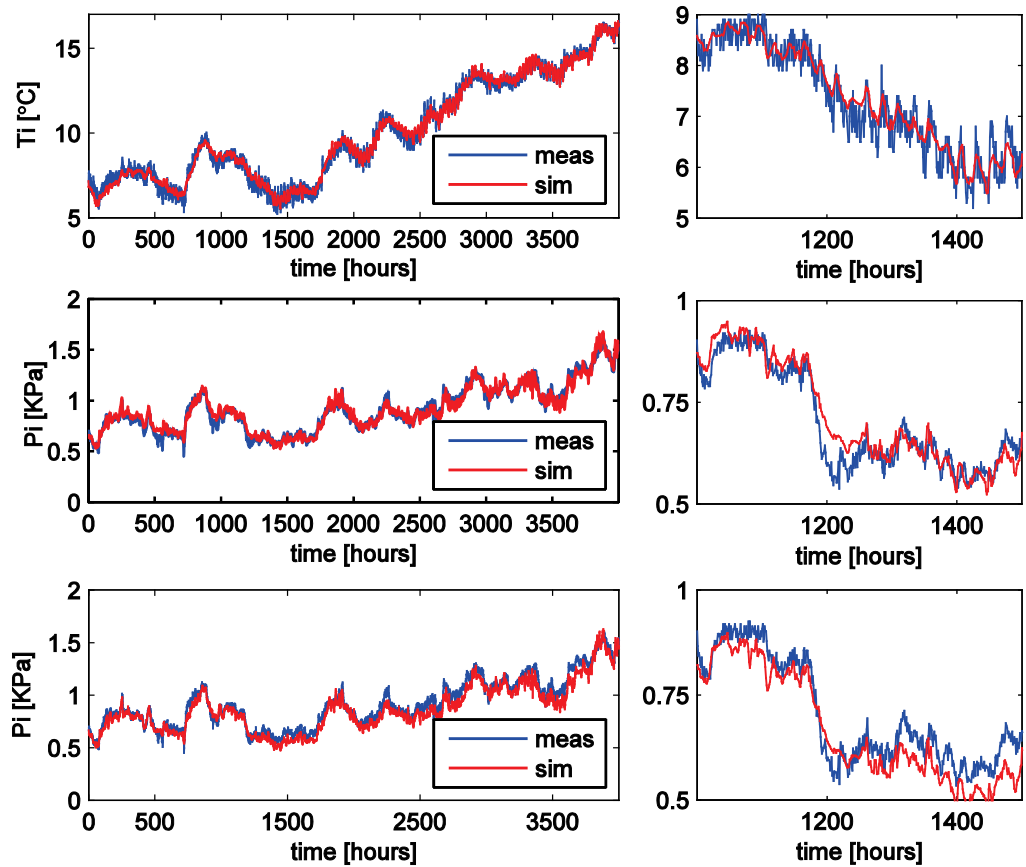

*Figure 4. The thermal model (top), hygric model with additional fixed vapour pressure node in model (middle) and hygric model without fixed pressure (bottom) fitted to washing room (This figure relies on color see digital version of the paper).*

*Castle of Gaasbeek - Scockaert chamber*
The Scockaert chamber is situated at the second floor in the castle of Gaasbeek (Belgium). It has windows which allow a significant amount of sun irradiation into the room. The room is richly decorated with antique furniture which adds to the thermal, and mostly, the hygric capacity. The difficulty here: the irregular disturbances by visitors. Because this effect can't be captured by a linear time invariant model, the Scockaert chamber is a good candidate for the models performance assessment. The result of the thermal model is shown in Figure 5 (top). The overall performance is good, but incidentally, some measured peaks are observed which are not reproduced by the model. These peaks might be a result of internal heat sources like visitors. The result of the hygric model (with fixed vapour pressure node) is shown in Figure 6 (bottom). The same observation holds as for the thermal result. The overall performance is good, but identical series of peaks are not reproduced. These peaks are due to inputs which are not included, e.g. moisture production by visitors.

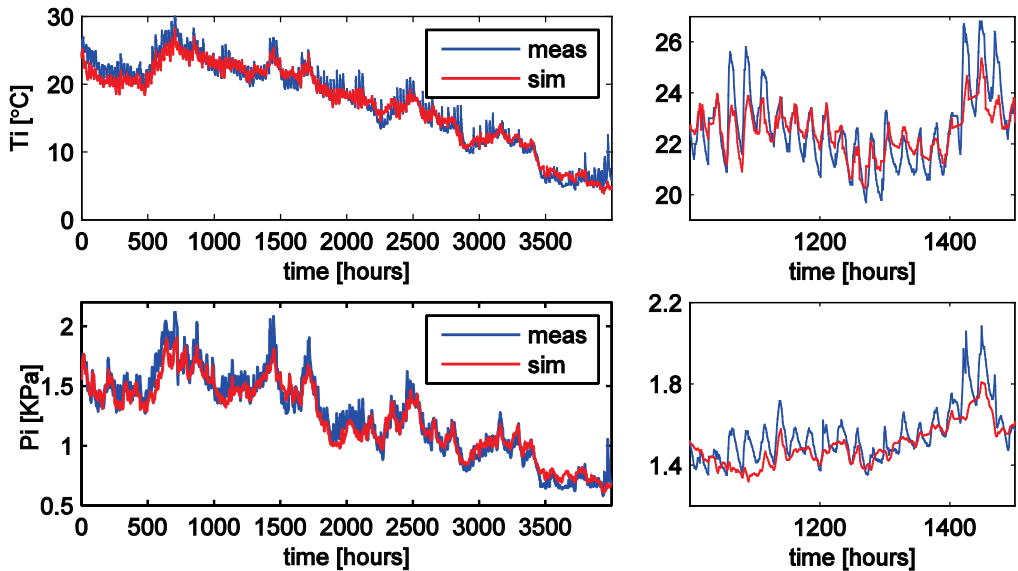

*Figure 5. The thermal model (top) and hygric model with additional fixed vapour pressure node (bottom) fitted to Scockaert chamber (This figure relies on color see digital version of the paper).*

*Castle of Gaasbeek - Gothic chamber*

The Gothic chamber is also situated at the second floor of the castle of Gaasbeek. The room is decorated richly with much furniture and decorations which contribute to the thermal, but mostly hygric capacity. Also this room is frequently visited by tourists. The performance of the thermal model is shown in Figure 6 (top). The incidental peaks are less intense compared to those in the Scockaert chamber, but are still visible. The detail at the right side of the figure shows that the model reproduces the measured signal accurately. The delicate quick fluctuations which are present in the measurements of the washing room are totally absent in the measurements of the Gothic chamber.

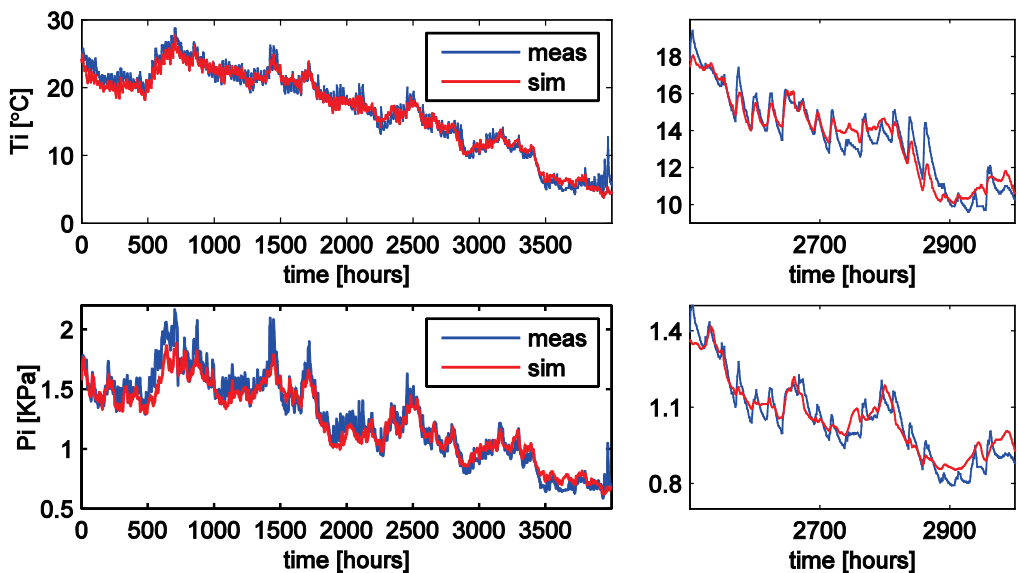

*Figure 6. The thermal model (top) and hygric model with additional fixed vapour pressure node (bottom) fitted to Gothic chamber (This figure relies on color see digital version of the paper).*

The hygric model performance is shown in Figure 6 (bottom). The overall performance is good, but small disturbances which are not reproduced are probably due to time-variant phenomena. At least, the phenomena which could change over the seasons turn out to be no source of error: think of possible use of sun blinds in summer.

*Castle Keukenhof - the loft*
The castle Keukenhof is situated a few kilometers from the Dutch coast. Therefore the building is exposed to a sea climate. The observed room is the loft which is situated directly underneath the roof. This room is specifically interesting because the overheating risk during summer. This situation will be a real test for the model since the influence of sun irradiation is significantly present, but windows are absent. The thermal model performance is visualized in Figure 7(top). The risk of overheating is clearly present and is reproduced good by the model. Although the sun irradiation is modeled by 4 input signals, each representing the sun irradiation on a vertical wall oriented respectively to the north, east, south and west, the sun irradiation on the roof and the resulting heat flux to the loft is simulated correctly. This significant heat flux resulting from sun irradiation results in an indoor temperature of reaching 35°C in summer. The detail at the right side of the figure shows how accurate the models output has been fitted to the measured temperature. The performance assessment of the hygric model is shown in Figure 7 (bottom). The hygric reproducibility is poor. Besides the indoor temperature, also the indoor moisture content (or vapour pressure) fluctuates heavily. These fluctuations good not be reproduced with the only given input of the outdoor vapour pressure. This casus is interesting for a follow-up research.

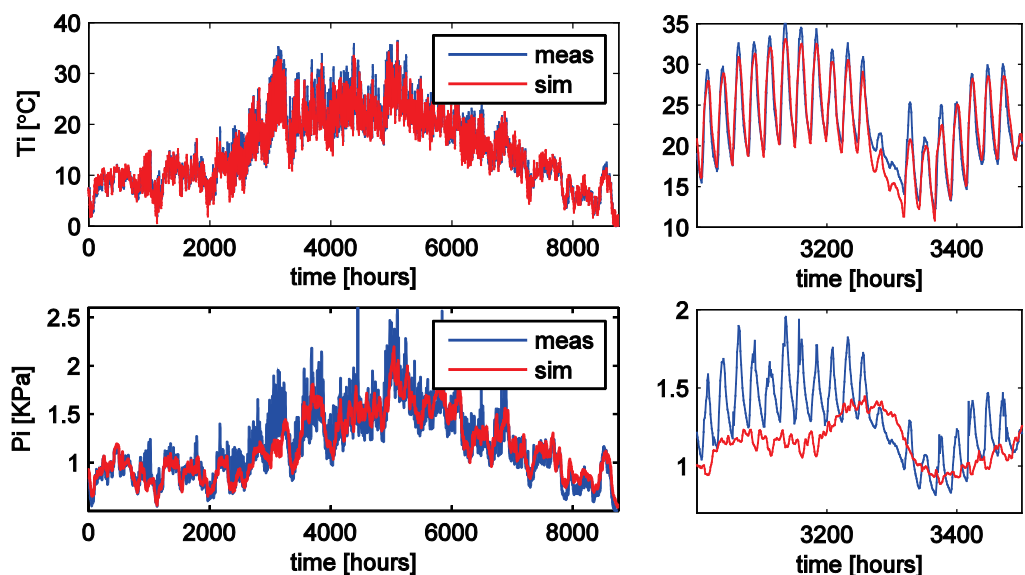

*Figure 7. thermal model (top) and hygric model with additional fixed vapour pressure node (bottom) fitted to Keukenhof-Loft (This figure relies on color see digital version of the paper).*

*St. Baafs cathedral – South transept*
The St. Baafs cathedral is situated in Gent (Belgium). The building consists of several parts, e.g. the choir, the entrance, the transepts. Most of the parts are openly connected. The building is totally free floating and has huge thermal mass and hygric mass.

The measurements to which the model has been fitted are performed by a sensor in the south transept. The south transept includes a huge window with colored glass which is oriented to the south resulting in a significant amount of incoming sun irradiation. The performance of the thermal model is shown in Figure 8 (top). The measured indoor temperature is reproduced very accurately, also visualized clearly in the detailed graph on the right side. Although there are visitors, the influence on the indoor climate seems to be small due to the large dimensions of the cathedral. The performance of hygric model is shown in Figure 8 (bottom). The measured signal is reproduced perfectly by the model. The same holds for the hygric part: although there are visitors, they hardly influence the indoor moisture content due to the vast space in the cathedral. Moreover, the quality of the measurements are high without signal noise.

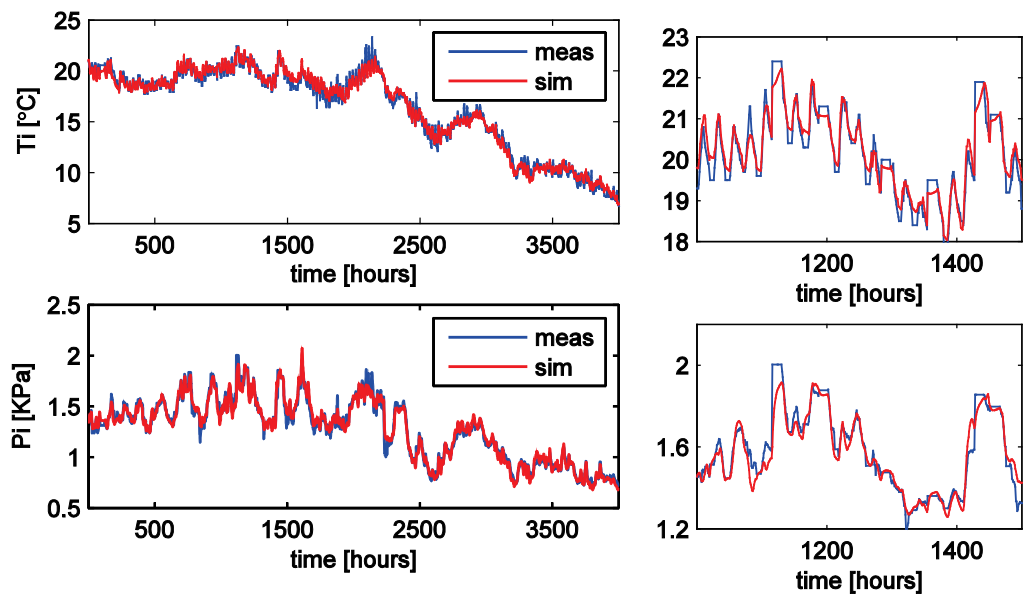

*Figure 8. The thermal model(top) and hygric model with additional fixed vapour pressure node (bottom) fitted to St. Baafs cathedral-South transept (This figure relies on color see digital version of the paper).*

**ASSESMENT OF THE METHODOLOGY**
This section deals with the performance assessment of the models and the used methods. The performance of the models relates to two things:
  i)   how accurately can the measured signal be reproduced
  ii)  how accurately can the parameters be identified (in validation section)
This implies a contradiction. Usually, the rule holds that a higher order model with more parameters yields a higher accuracy. On the other hand, more parameters result in a higher uncertainty in parameter identification (Ljung, 1999).
To be able to compare and rate how accurately the different models can reproduce a measured temperature or vapour pressure, three performance criteria are used: the MSE (Mean Squared Error), MAE (Mean Absolute Error) and FIT (goodness of FIT).
The MSE is calculated according to,

$$MSE = \frac{1}{N}\sum_{k=1}^{N}(y' - y)^2$$

where $y'$ is the measured signal and $y$ is the simulated signal.

The MAE is calculated according to,

$$MAE = \frac{1}{N}\sum_{k=1}^{N}|y' - y|$$

The Goodness of Fit is calculated according to,

$$FIT = 100 \cdot \left(1 - \frac{norm(y' - y)}{norm(y' - \bar{y}')}\right)$$

$norm(y)$ is the Euclidean length of the vector $y$, also known as the magnitude. The above equation therefore calculates in the numerator the magnitude of the error between measured and simulated signal. This is divided by the denominator, calculating how much the measured signal fluctuates around its mean. Consequently, the Goodness of Fit criterion is robust with respect to the fluctuation level of the signal. All three are used independently in several studies, e.g. (Boaventura Cunha, et al., 1997) (Crabb, et al., 1987) (Frausto, et al., 2003). However, (Mustafaraj, et al., 2010) stresses the strength of using the three together. For example, the MSE gives more weight to bigger errors. Consequently, it is a good criterion to express the amount of big errors. The MAE expresses the overall mean error.

*Table 2. results of fit summarized in three performance criteria (MSE, MAE & FIT).*

| building | room | thermal model 4a | | | hygric model 3 | | |
|---|---|---|---|---|---|---|---|
| | | MSE [°C$^2$] | MAE [°C] | FIT [%] | MSE [Pa$^2$] | MAE [Pa] | FIT [%] |
| Amerongen | washing room* | 0.09 | 0.24 | 91 | 4938 | 58 | 72 |
| | washing room | | | | 2468 | 39 | 81 |
| Gaasbeek | Scockaert chamber | 1.03 | 0.73 | 84 | 5369 | 57 | 77 |
| | Gothic chamber | 0.66 | 0.59 | 87 | 5316 | 56 | 78 |
| Keukenhof | loft | 1.33 | 0.88 | 84 | 23785 | 113 | 58 |
| St. Baafs cathedral | south transept | 0.17 | 0.32 | 91 | 1870 | 32 | 86 |
| | south transept** | 0.74 | 0.69 | 81 | | | |
| museum Gevangen-poort | dungeon of pain | 0.30 | 0.41 | 87 | 3686 | 47 | 84 |
| | iron chamber | 0.10 | 0.26 | 82 | 462 | 18 | 85 |
| | knight chamber | 0.34 | 0.45 | 88 | 3710 | 47 | 84 |
| | stock loft | 1.41 | 0.96 | 81 | 6954 | 65 | 78 |

*model without node for fixed vapour pressure

**no solar input

**CONCLUSIONS**

It is concluded that this approach is very promising for obtaining physical models and parameters of indoor climate responses. Furthermore by using state space models, the simulation duration time of a 100 year hourly based period take less than a second on an ordinary computer.